\documentclass{article}

\usepackage{arxiv}

\usepackage[utf8]{inputenc} 
\usepackage[T1]{fontenc}    
\usepackage{hyperref}       
\usepackage{url}            
\usepackage{booktabs}       
\usepackage{amsfonts}       
\usepackage{nicefrac}       
\usepackage{microtype}      
\usepackage{lipsum}

\usepackage{abstract}

\usepackage{graphicx}
\graphicspath{ {./images/} }
\usepackage{amsmath}
\usepackage{graphicx}
\usepackage{xcolor}
\usepackage{subcaption}
\begin{document}
\title{asLLR:LLM based Leads Ranking in Auto Sales}

\author{
    Yin Sun\\
    Li Auto Inc.\\
    sunyin@lixiang.com\\
    \And
    Yiwen Liu \\
    Li Auto Inc.\\
    liuyiwen@lixiang.com\\
    \And
    Junjie Song \\
    Li Auto Inc.\\
    sjj50179@163.com \\
    \And
    Chenyu Zhang\\
    Li Auto Inc.\\
    cyzhang57@outlook.com\\
    \And
    Xinyuan Zhang \\
    Li Auto Inc.\\
    xyzhang607@163.com \\
    \And
    Lingjie Liu\\
    Li Auto Inc.\\
    lingjie26@126.com \\
    \And
    Siqi Chen\\
    Li Auto Inc.\\
    Chan\_4\_7@sjtu.edu.cn \\
    \And
    Yuji Cao* \\
    Li Auto Inc.\\
    caoyuji@lixiang.com
}
\maketitle
\begin{abstract}

In commercial auto sales systems, high-quality lead scoring establishes the priority of sales efforts and is critical to system efficiency. Since CRM (Customer Relationship Management) system contains many textual interaction features between sales and customers,
traditional techniques such as Click Through Rate (CTR) prediction struggle with processing the complex information inherent in natural language features, which limits their effectiveness in sales lead ranking. Bridging this gap is critical for improving business intelligence and decision-making. Recently, the emergence of large language models (LLMs) has opened new avenues for improving recommendation systems,
this study introduces asLLR (LLM-based Leads Ranking in Auto Sales), which integrates CTR loss and Question Answering (QA) loss within a decoder-only large language model architecture. This integration enables the simultaneous modeling of both tabular and natural language features.
To verify the efficacy of asLLR, we constructed an innovative dataset derived from the customer lead pool of a prominent new energy vehicle brand, with 300,000 training samples and 40,000 testing samples. Our experimental results demonstrate that asLLR effectively models intricate patterns in commercial datasets, achieving the AUC of \textbf{0.8127}, surpassing traditional CTR estimation methods by \textbf{0.0231}. Moreover, asLLR enhances CTR models when used for extracting text features by \textbf{0.0058}. In real-world sales scenarios, after rigorous online A / B tests, asLLR increased sales volume by about \textbf{9.5}\% compared to the traditional method.
\end{abstract}

{\textbf{Code} -- \url{https://github.com/alg-znsy-li/as_llr}}

\section{Introduction}

In the automotive sector, vehicle sales constitute a pivotal stage for revenue generation. Research has shown that AI-driven sales systems can enhance customer conversion rates by over 20\% compared to more traditional sales approaches\cite{fischer2022artificial,alavi2024salesperson,fehrenbach2025artificial}. 
Given the status of automobiles as high-value consumer goods, customers often engage in comparative analysis between different manufacturers, significantly motivating many clients to explore online sales platforms. 
In this digital engagement phase, sales specialists maintain regular follow-ups with potential leads ( In the sales domain, "leads" refer to potential customers or prospective business opportunities. These leads are typically acquired through various marketing initiatives. Each lead signifies a potential sales opportunity, warranting focused attention and follow-up by the sales team ).
The diversity of customer purchasing intentions requires sales professionals to focus on acquiring high-quality leads. This imperative requires astute prioritization of business leads within the technical framework of automotive sales\cite{sf}. Fundamentally, this challenge pertains to resolving the issue of accurately estimating the click-through rate (CTR) to assess the quality of sales leads.

Current mainstream methods for Click-Through Rate (CTR) prediction predominantly rely on deep learning architectures or statistical learning paradigms. Deep neural networks have achieved remarkable success in this domain\cite{zhang2016deep},  while existing DNN-based approaches demonstrate competence in processing structured tabular data\cite{zhang2024wukong}, they exhibit inherent constraints in capturing nuanced semantic patterns within unstructured natural language features. This represents a significant drawback given that textual information often contains decisive indicators that influencing consumer purchasing decisions.

Therefore, our primary research focus on how to simultaneously model both tabular and textual features. 
Due to the emergence of large language models (LLMs) presents new opportunities through their exceptional natural language understanding capabilities and few-shot learning capabilities\cite{lmaskb,gpt3analysis,agent}, we introduce asLLR, a LLM based Leads Ranking model in auto sales. This model leverages the strengths of large language models to effectively manage both tabular and textual data features. Moreover, we integrate CTR loss and QA loss to improve the model's proficiency in interpreting input data. Empirical evaluations reveal that asLLR achieves an impressive AUC of 0.8127, markedly outperforming the current leading CTR estimation models. Moreover, stringent online A/B testing indicates that the asLLR model enhances business order conversion rates by 9.5\% over existing online methods, underscoring its efficacy in real-world scenarios.

Moreover, we have identified a notable gap in the existing academic literature regarding the evaluation of lead quality in the auto sales sector. Therefore, we created a leads quality assessment dataset specifically designed for automotive retail sector, utilizing authentic data from the retail system of a particular electric vehicle company. This dataset comprises 300,000 training samples and 40,000 test samples. Each sample encompasses four categories of 31 tabular features, along with a communication record detailing the interactions between the sales expert and the customer. 

We demonstrated the effectiveness of the asLLR model through real online A/B testing. However, we found that in the real world, excessively long textual features can lead to a decline in the model's performance.
To address this challenge, we integrated a text summarization module into asLLR. This module performs knowledge compression on the input data, effectively alleviating potential declines in model performance when dealing with extended textual content. Experimental results indicate that the incorporation of a text summarization module significantly enhances the model's performance in handling long text inputs. 

In summary, the main contributions of our paper are:
\begin{itemize}

\item We introduce asLLR, a LLM based Leads Ranking model in Auto Sales. This model effectively manage both tabular and textual data features, improving its performance in addressing the issue of lead evaluation in the automotive sales sector.
\item We created a leads quality assessment dataset specifically designed for automotive retail sector. In assembling this dataset, we meticulously differentiated the temporal sequences of the training and testing sets to uphold the scientific integrity and robustness of the model evaluation process.
\item  We integrated a text summarization module into asLLR, improving the model's performance in handling long text inputs.  

\end{itemize}
\section{Related Work}

\subsection{Click-Through-Rate Prediction}
Within the industry, there exist numerous solutions to ranking problems. Acknowledging the clear superiority of deep learning methods over traditional statistical machine learning approaches \cite{cheng2016wide}, our subsequent discussion will focus exclusively on deep learning techniques. Presently, three primary approaches are employed to address ranking problems using deep learning: (1) vector-based methods, (2) bit-based methods, and (3) graph neural network-based methods. 
Vector-based methods, as exemplified by models like DeepFM \cite{deepfm}, normalize all features into vectors of uniform size, thereby allowing the exploration of second-order feature interactions. This approach combines deep and wide networks to enhance memory and generalization capabilities. However, it is theoretically limited in its ability to model feature interactions beyond second-order. To overcome this limitation, bit-based methods have been developed. Unlike vector-based methods, bit-based approaches do not require uniform vector alignment of all features \cite{dcn,dcnm,dcnv2}. This flexibility allows for effective modeling of higher-order feature interactions, with DCN \cite{dcn} serving as a notable example.
In addition to these, recent advancements have introduced graph neural network-based methods. These techniques conceptualize first-order features as nodes within a graph, with feature interactions represented as edges. Algorithms are employed to manage the generation and selection of these edges for modeling feature interactions \cite{graphfm}. To address interactions of higher orders (beyond second-order), the notion of hypergraphs is utilized, giving rise to hypergraph neural networks \cite{hirs}. Furthermore, other innovative approaches, such as attention-based technologies like AutoInt, have emerged in the study of ranking problems, representing methodologies that do not fit neatly into the aforementioned categories.
\subsection{Conversation Summary}

Text summarization \cite{el2021automatic,nallapati2016abstractive,shi2021neural,fabbri2019multi,li2019abstractive,deyoung2021ms2,nallapati2017summarunner,narayan2018ranking,zhong2020extractive,zhang2020pegasus} aims to condense a set of texts into a concise summary that retains essential information. 
In the early stages of text summarization, rule-based extractive methods predominated, including techniques like Lead-3 and TextRank \cite{mihalcea2004textrank}. With the emergence of neural network technology, the focus has increasingly shifted towards abstractive summarization using deep learning-based Seq2Seq models. BERT \cite{devlin2018bert} emerged as a prominent method for abstractive summarization, achieving commendable outcomes.

In recent years, with the rapid advancement of LLM technology, more researchers have chosen to utilize large models for text summarization tasks. These models have become a dominant solution due to their superior instruction-following ability and semantic understanding capabilities. Studies \cite{pu2023summarization} indicate that the text summarization abilities of large models surpass all previous mainstream techniques and even exceed the average quality of human-generated summaries. In this study, we focus on using large language models as our primary approach to enhancing the informational density of long textual inputs. By summarizing lengthy dialogues with low informational density, we aim to reduce the input length while preserving a high level of information, thus improving the signal-to-noise ratio of the data.

\section{Method}

\begin{figure*}[h]
\centering
\centerline{\includegraphics[width=0.70\linewidth]{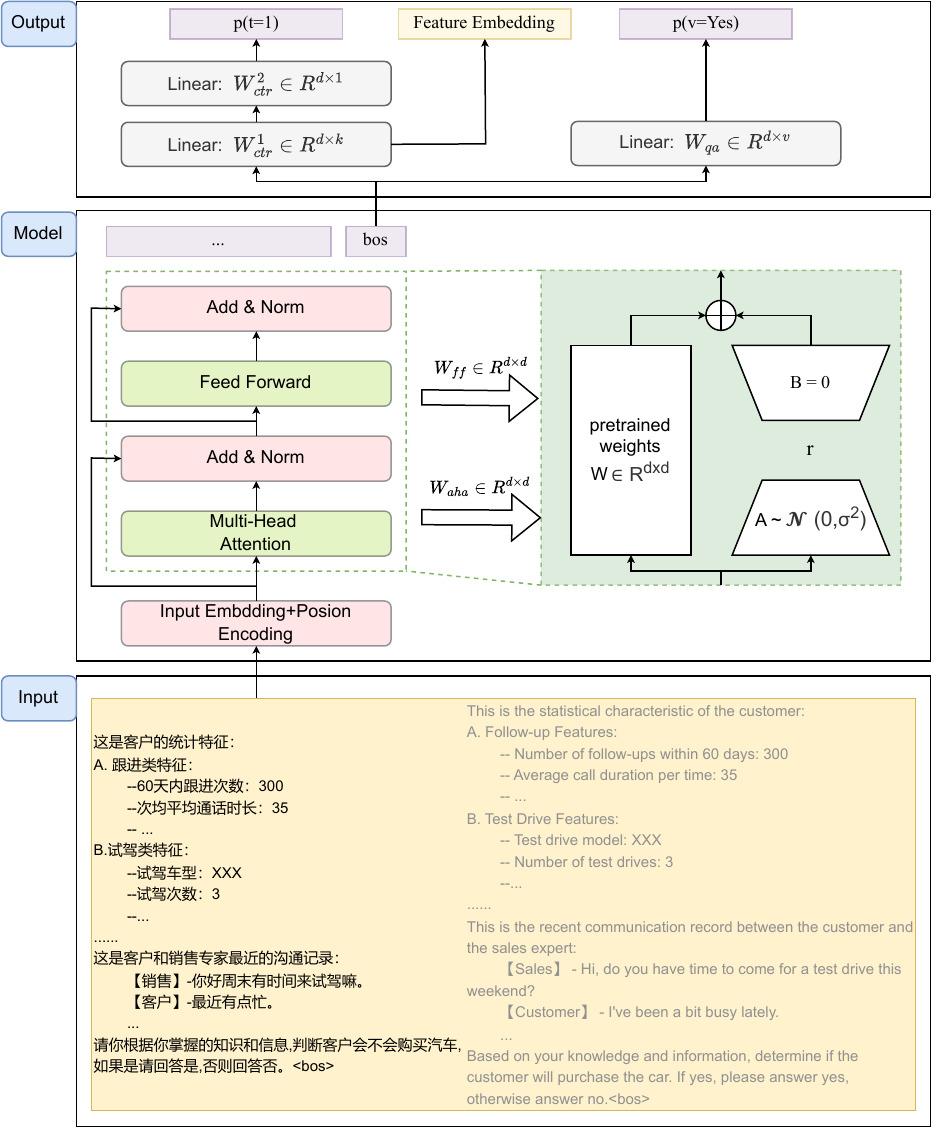}}
\caption{Overview of asLLR framework.}
\label{abstract_asllr}
\end{figure*}

\subsection{Problem Statement}
The asLLR model is meticulously crafted to tackle the intricate challenge of evaluating the quality of a significant volume of customer leads, which are curated by sales experts. We conceptualize this issue as a Learning-to-Rank problem, incorporating distinct operational modifications, and redefine it as the task of estimating purchase probability. In this context, we adopt a point-wise approach from ranking learning methodology. This choice is supported by two primary considerations: firstly, the point-wise approach is widely adopted in the contemporary mainstream models for CTR prediction \cite{inproceedings}; secondly, our dataset construction leverages natural outcome labels that are intrinsically binary, characterized by only two possible outcomes: purchase and non-purchase.

As previously described, the model takes two types of inputs: tabular features 
$T$ and text features $L$. For a customer $i$ the tabular features  $T^i = \{(n_j:t_j^i)|j=1,2,...,t\}$ represent statistical characterizations of the customer, where $t_j^i$ denotes a specific tabular feature of this customer, such as average call duration. Similarly, for this customer, the text features $L^i=\{l_j^i|j=1,2,...,l\}$ characterize the communication process between the customer and the sales expert, where each $l_j^i$
 represents a single character in the communication process. For convenience, we select the most recent $l$ length of communication records for text features. The customer sample is ultimately transformed into training labels $\{y^i|i=1,2,...,|D|\}$, $y_i=1$ means that the user purchased the car, while a value of $y_i=0$ indicates that the user did not make a purchase. This is formalized as follows:

\begin{equation}
    D=\{ [(T^i,L^{i});y^i]|i=1,2,\cdots,|D| \}.
\end{equation}

Assuming that our model parameters are $\theta =\{\alpha,\beta\}$, where $\alpha = \{      W_{ff}^i;W_{mha}^i|i=1,2,...,N\}$ represents the parameters of the base LLM  component, fine-tuned using the LoRA ( Low-Rank Adaptation)\cite{hu2022lora} technique, and $\beta = \{W_{ctr}^{1,2},W_{qa}\}$are the task adaptation parameters. Mathematically, this problem can be expressed as the following function fitting problem:

\begin{equation}
    p(y^i=1)=f(T^{i},L^{i};\theta).
\end{equation}

\begin{figure}[h]
    \centering
    \begin{subfigure}{0.9\textwidth}
        \centering
        \includegraphics[width=0.6\textwidth]{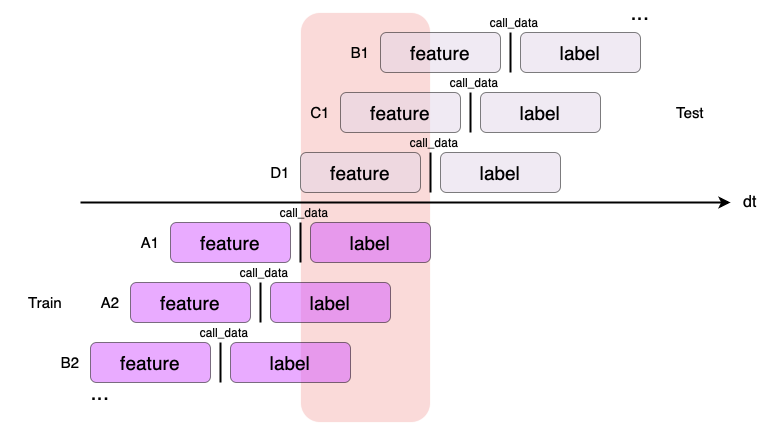}
        \caption{Method for Training and Test Set Partitioning to Prevent Data Leakage.\label{fig:image_data_split}}
    \end{subfigure}
    \vspace{0.5cm}
    \begin{subfigure}{0.45\textwidth}
        \centering
        \includegraphics[width=\textwidth]{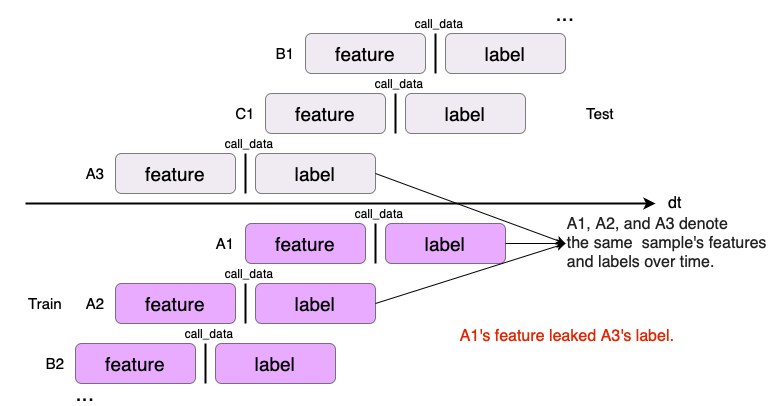}
        \caption{The Example of Feature Leakage.\label{fig:image_fea_leak}}
    \end{subfigure}
    \hfill
    \begin{subfigure}{0.45\textwidth}
        \centering
        \includegraphics[width=\textwidth]{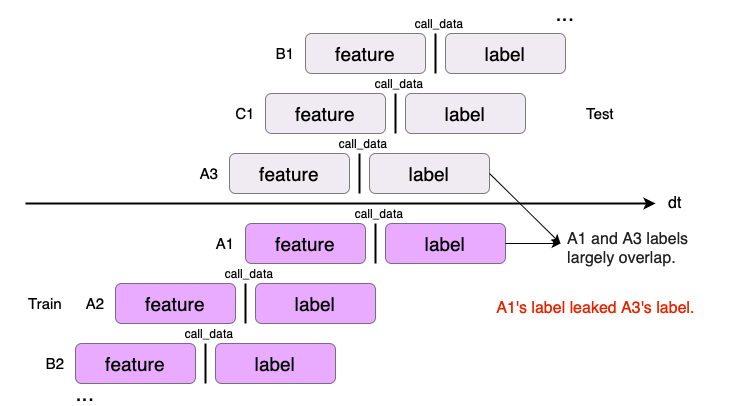}
        \caption{The Example of Label Leakage.\label{fig:image_label_leak}}
    \end{subfigure}
    \caption{Methodology for Dataset Construction.}
    \label{fig:threeimages}
\end{figure}

In crafting the model, we intentionally refrained from making intrusive alterations to the LLM models. This choice was informed by the rapid pace of advancements and frequent updates in LLM model development, which are marked by swift enhancements in performance. By opting for non-intrusive modifications, we ensure the flexibility to readily replace the underlying models, facilitating rapid iteration and adaptation to evolving technologies.

The asLLR model, as illustrated in Figure \ref{abstract_asllr}, is composed of three layers: the input layer, the model layer, and the output layer. The input layer is responsible for constructing prompts from features, the model layer consists primarily of transformer decoder blocks, and the output layer has three heads. The output heads on the sides are used for constructing the CTR (Click-Through Rate) loss and QA (Question Answering) loss, as previously mentioned. For a fair comparison, the middle output head of the output layer is used to generate an embedding representation of the text features. This embedding will serve as a feature in traditional CTR models during subsequent experiments, although at this point, the input layer does not include tabular features. When evaluating the asLLR model, the middle head of the output layer is not utilized.

The model's input layer processes tabular and text features into a unified natural language representation of customer descriptors. This section consists of two main components: the feature specification part and the inference prompt part. The feature specification part describes the customer's behavior characteristics and communication records in a hierarchical list format. In the inference prompt part, we define the model output format and provide an inference start prompt symbol $<bos>$.
The model layer is primarily composed of the Transformer's decoder blocks, which are fine-tuned using LoRA technology. For the $i$th sample customer, the model layer yields the representation $h^{i}_{bos}$ corresponding to the inference start prompt symbol $<bos>$. The output layer includes three linear layers, which are responsible for:

\begin{align}
    h_{\text{ctr1}}^i &= h_{\text{bos}}^i W_{\text{ctr}}^1 ,\\
    h_{\text{ctr}}^i &= h_{\text{ctr1}}^i W_{\text{ctr}}^2 ,\\
    h_{\text{qa}}^i &= h_{\text{bos}}^i W_{\text{qa}}.
\end{align}

For the asLLR model, there are two types of losses: the CTR loss and the QA (Question Answering) loss. Regarding the CTR loss and the QA loss, we have:

\begin{align}
\mathcal{L}_{\text{CTR}} &= -\frac{1}{|D|} \sum_{i=1}^{|D|} [y^i \log(\text{sigmoid}(\tilde{h}_{\text{ctr}}^i)) + \\&\qquad(1 - y^i) \log(1 - \text{sigmoid}(\tilde{h}_{\text{ctr}}^i))].\\
\mathcal{L}_{\text{QA}} &= -\frac{1}{|D|} \sum_{i=1}^{|D|} [y^i \log(\text{softmax}(\tilde{h}_{\text{qa}}^i)[v_{\text{yes}}]) + \\&\qquad(1 - y^i) \log(\text{softmax}(\tilde{h}_{\text{qa}}^i)[v_{\text{no}}])].
\end{align}
Ultimately, these two losses are combined during the training process:

\begin{align}
\mathcal{L} = \mathcal{L}_{\text{CTR}} + \mathcal{L}_{\text{QA}}.
\end{align}

The rationale for constructing two distinct loss functions stems from our experimental findings, which showed that employing only the CTR loss led to severe overfitting in the asLLR model. However, the introduction of the QA loss mitigated this overfitting issue. We hypothesize that the QA loss, being rooted in the domain of natural language processing, enables more effective utilization of the knowledge embedded within the LLM.
 So we use different learning rates to fit the parameters $\alpha$ (LLM layers) and $\beta$ (Output layers): a lower learning rate for $\alpha$ and a higher learning rate for $\beta$.

\section{Dataset}

Due to the lack of publicly available academic datasets in auto sales, we collected approximately 300,000 customer leads from a certain new energy vehicle retail system as a training set and 40,000 customer leads as a test set. Each sample represents a customer; the label is 1 if the customer eventually purchased a vehicle and 0 otherwise. The training and test sets do not overlap temporally, ensuring that the model does not overfit time-based confounding factors. The dataset includes two types of features: tabular features, which record various statistical behavioral features of customers, and textual features, which document the communication process between sales and customers.
Specifically, each sample comprises 31 types of tabular features classified into four categories, alongside a record of the most recent communication between the sales and the customer. We evaluated the effective coverage rate of each tabular feature and observed that more than 93\% of these features exhibited an effective coverage rate exceeding 90\%. The communication records between sales experts and customers are predominantly derived from telephone conversations, which we converted into text sequences using Automatic Speech Recognition (ASR) technology. The ratio of positive to negative samples in both the training and test datasets is approximately 1.45\%.
To our knowledge, this dataset represents the first large-scale model training and testing dataset constructed for the academic field of automotive sales lead ranking. 

In conventional dataset construction, the data is randomly partitioned into training and test sets based on IDs. However, this approach might result in data leakage, meaning that the model uses information from the future and is inconsistent with the paradigm of online inference. Although users in the training and test sets differ, patterns or features learned by the model might be shared or interdependent among different users. This implies that the data present in the training set could indirectly affect the model predictions. To address this issue, we have implemented a solution to ensure that the time range of the training set is entirely prior to that of the test set. The specific dataset partitioning scheme is illustrated in Figure \ref{fig:image_data_split}, where the red area represents the shortest date interval between the training and test sets, corresponding to the date range of the labels. This partitioning method ensures the consistency of the training and inference paradigms both online and offline, effectively resolving issues related to feature leakage and label leakage.

Feature leakage refers to inappropriate dependencies between features and the target variable, such as directly using some form of test data labels as features. An example of this is shown in Figure \ref{fig:image_fea_leak}. Label leakage occurs when label information during the model training phase affects model performance in the validation phase, for instance, when there is correlation between the labels of certain training and test samples, and overlaps exist in their label ranges. An example of this is shown in Figure \ref{fig:image_label_leak}.

\section{Experiments}

To evaluate the performance of the model, we employ the most common metric in CTR prediction: Area Under Curve (AUC). AUC is a metric that is independent of the ratio between positive and negative samples and is highly effective in capturing the model's ranking ability. A higher AUC indicates a greater probability that the model ranks high-quality leads in the top positions.

To comprehensively assess the superiority of our proposed technique, we designed a relatively rigorous set of baseline models. We selected six classic baseline models for performance evaluation. First, we selected representative networks from vector-based CTR models, such as DeepFM\cite{deepfm}, and representative networks from bit-based CTR models, such as DCN\cite{dcn}, along with other significant networks for reference. Second, we chose several classic ranking models (W\&D\cite{cheng2016wide}) and models based on the transformer architecture (AutoInt\cite{lindell2021autoint}).
This comparison is intended to evaluate whether our proposed method, asLLR, exhibits technical superiority over traditional deep learning methods. Furthermore, we used pre-training techniques to convert text features into 128-dimensional vectors for input into the aforementioned CTR models, which enhances the fairness of model comparisons. Additionally, throughout the comparison process, important training parameters such as batch size, epoch, and optimizer are kept consistent unless otherwise specified. The configuration of baseline models and training parameters is detailed in Table \ref{tab:tab1}.
To prevent overfitting, we used LoRA technology ( $rank  = 16, \alpha=1$) to fine-tune the base model and employed a lower learning rate. 
\begin{table}[hb]
\vspace {-2.5mm}
\begin{center}
\caption {The configuration of the baseline model and training parameters.}
\resizebox{0.7\columnwidth}{!}{
\begin{tabular}{ll}
\toprule
Model & configuration \\
\hline
W\&D & \{embedding\_dim:8\}
\\
DeepFM & \{embedding\_dim:8\}
 \\
xDeepFM & \{embedding\_dim:8, cin\_layer\_size:[256,128]\} 
 \\
DCN&\{embedding\_dim:8, cross\_num:2\}
 \\
DCN-M & \{embedding\_dim:8, cross\_num:2\}
 \\
AutoInt & \{embedding\_dim:8, att\_layer\_num:3, att\_head\_num=2\}
 \\
\bottomrule
\end{tabular}
}
\label{tab:tab1}
\end{center}
\end{table}
\begin{table}[htbp]

\vspace {-2.5mm}
\begin{center}
\caption {Main Result. $+h_{ctr1}$ represents the model's use of the asLLR output layer's feature embedding as input.}
\resizebox{0.7\columnwidth}{!}{
\begin{tabular}{l|c|l|c}
\toprule
Model &  $+h_{ctr1}$&w empty samples & w/o empty samples\\
\hline
W\&D &No& 0.7860 & -
\\
DeepFM &No& 0.7917 & - 
\\
xDeepFM &No& 0.7808 & - 
\\
DCN &No&  0.7862 & - 
\\
DCN-M &No&  0.7900 & - 
\\ 
AutoInt &No&  0.7896 & - 
\\ 
\hline
W\&D&Yes& 0.7951(+0.0091) & - 
\\
DeepFM &Yes& 0.7976(+0.0059) & - 
\\
xDeepFM &Yes& 0.7895(+0.0087) & - 
\\
DCN &Yes&  0.7911(+0.0049) & - 
\\
DCN-M&Yes& 0.7911(+0.0011) & - 
\\
AutoInt&Yes & 0.7950(+0.0054) & - 
\\
\hline
asLLR +ctr &-& 0.7921 & 0.8116 
\\
asLLR +ctr +qa &-& $\mathbf{0.8081} $ & $\mathbf{0.8127}$
\\
\bottomrule
\end{tabular}
}
\label{tab:tab2}
\end{center}
\end{table}

\subsection{Main Results}
In our dataset, we noted that some leads lacked interaction with sales experts, leading to instances of empty textual features for these samples. This absence of data could potentially affect the model's performance. Consequently, we considered this factor in the design of our baseline. In our primary results, we conducted comparative analyses across two dimensions: the inclusion or exclusion of empty textual features during the evaluation process.

From Table \ref{tab:tab2}, it is evident that under the same conditions, our proposed asLLR model significantly outperforms existing representative CTR models. On one hand, we observe that merely incorporating the CTR loss from the numerical domain allows the asLLR model to achieve performance comparable to traditional CTR models in terms of AUC. On the other hand, if we simultaneously add the numerical domain CTR loss and the textual domain QA loss, the AUC of the asLLR model significantly surpasses that of traditional CTR models ($0.7917$ vs. $0.8081$). Additionally, to eliminate the effect of the proportion of empty textual samples on the model, we evaluated the model's performance after removing empty textual samples. Approximately $15\%$ of the samples in our dataset are empty textual samples, and during evaluation, these samples are removed from both the training and test sets. We found that upon removal of empty samples, due to the asLLR model's superior capability in modeling natural language features, its performance further improves (0.8081 vs. 0.8127), which is very consistent with our intuition.

Additionally, our proposed asLLR model serves as an effective natural language feature extractor. From Table \ref{tab:tab2}, we observe that incorporating the natural language embedding features $h^i_{ctr1}$ extracted by asLLR as 128-dimensional numerical features directly into the traditional CTR model can significantly enhance its performance. We notice an average AUC improvement of approximately 0.005. 

To gain a deeper understanding of the performance of the asLLR model, we conducted experiments under the optimal training setup ($\{batch\_size:256, drop\_out:0.5, rank:16, alpha:1, epoch:1\}$) to evaluate the impact of different base models, varying model parameter scales, training data sizes, and multiple tasks across different feature categories. It is important to note that we discovered using approximately 5\% of the training steps for learning rate warm-up significantly enhances the performance of the asLLR model. Consequently, the remaining discussion in this section focuses on experimental results under this learning rate strategy. Additionally, unless otherwise specified, all subsequent results pertain to the performance of the model in a multi-task setting.
\begin{figure}[htbp]
\centering
\centerline{\includegraphics[width=0.7\linewidth]{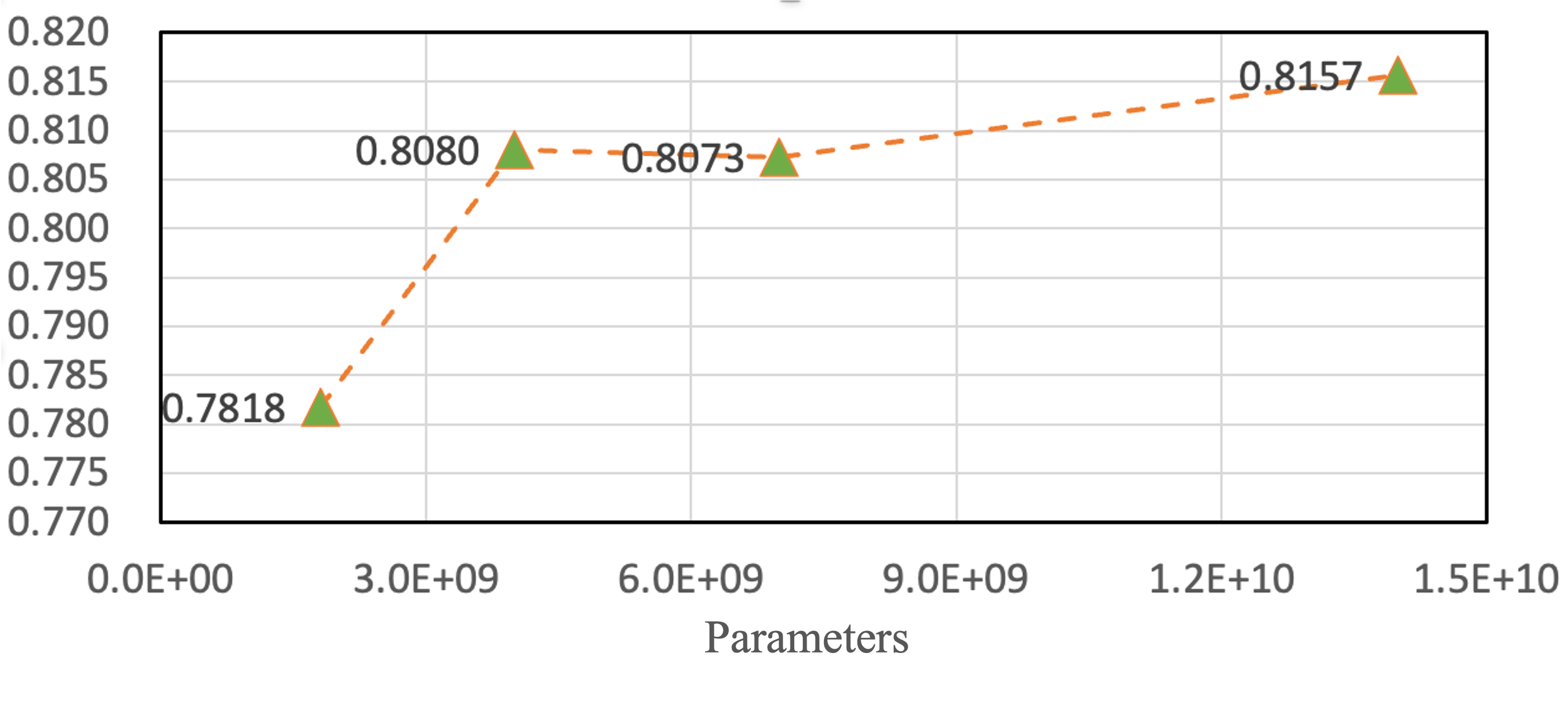}}
\caption{\quad  The Impact of Model Parameter Size on Performance.}
\label{fig:f2_0}
\end{figure}
\begin{figure}[h]
\centering
\centerline{\includegraphics[width=0.7\linewidth]{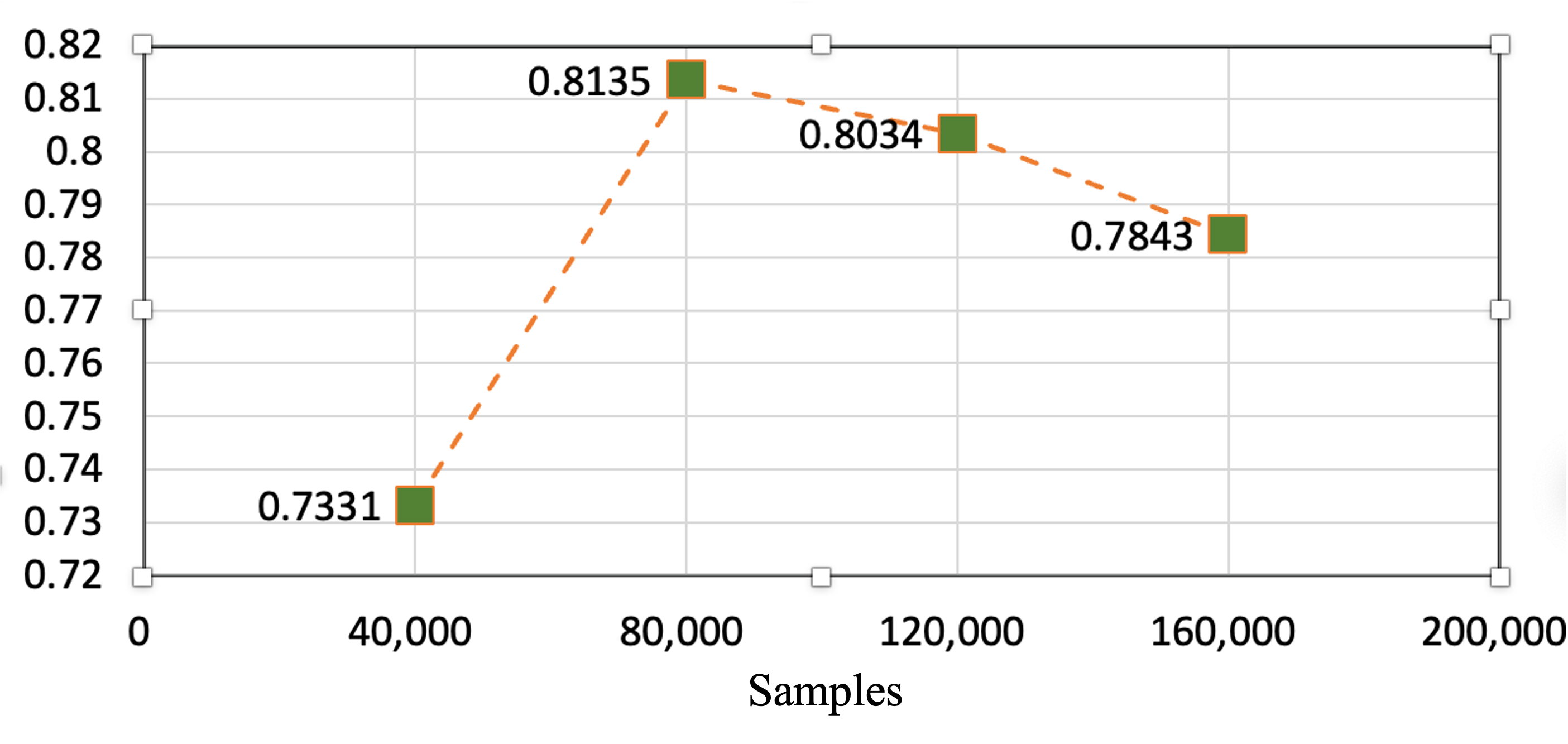}}
\caption{\quad  The Influence of Training Data Volume on Model Performance.}
\label{fig:f2}
\end{figure}

\subsection{Scaling Law for asLLR}

In the field of natural language processing, existing research suggests that large language models (LLMs) exhibit a scaling law\cite{kaplan2020scaling} indicating relationships between model parameter size, training data volume, and model performance: 1) The more model parameters, the better the model performance; 2) The more training data, the better the model performance. Given that our foundational model also employs LLMs, we investigate whether the asLLR technique demonstrates a similar scaling law in our context. To explore this, we examine the relationship between the effectiveness of the asLLR technique and both training data volume and model parameter size. We selected Qwen1.5 as the base model for evaluation. For Qwen1.5, we evaluated four configurations: 1.8B, 4B, 7B, and 14B parameters. As shown in Figure \ref{fig:f2_0}, it is clear that as the parameter size increases, model performance improves. However, once the parameter count surpasses a certain threshold, the diminishing returns become apparent, reflecting the boundary effect in performance enhancement.

From Figure \ref{fig:f2}, we observe that the performance of the asLLR model does not consistently improve with an increase in training data. Instead, it initially enhances and then gradually deteriorates. We hypothesize two potential reasons for the absence of a clear scaling law in the asLLR model. First, our dataset was derived from ASR outputs with relatively low character accuracy, which might disrupt the linguistic symbol logic learned by the base model as the training data increases. Second, our data might be overly concentrated, leading to catastrophic forgetting\cite{kirkpatrick2017overcoming} in the base model. A scientific analysis of the underlying causes of this phenomenon is a vital direction for our future research: we plan to adopt more advanced ASR models to refine our text data and also perform certain model modifications.

\begin{figure}[ht]
    \centering
    \begin{subfigure}[t]{0.45\textwidth}
        \centering
        \includegraphics[width=\linewidth]{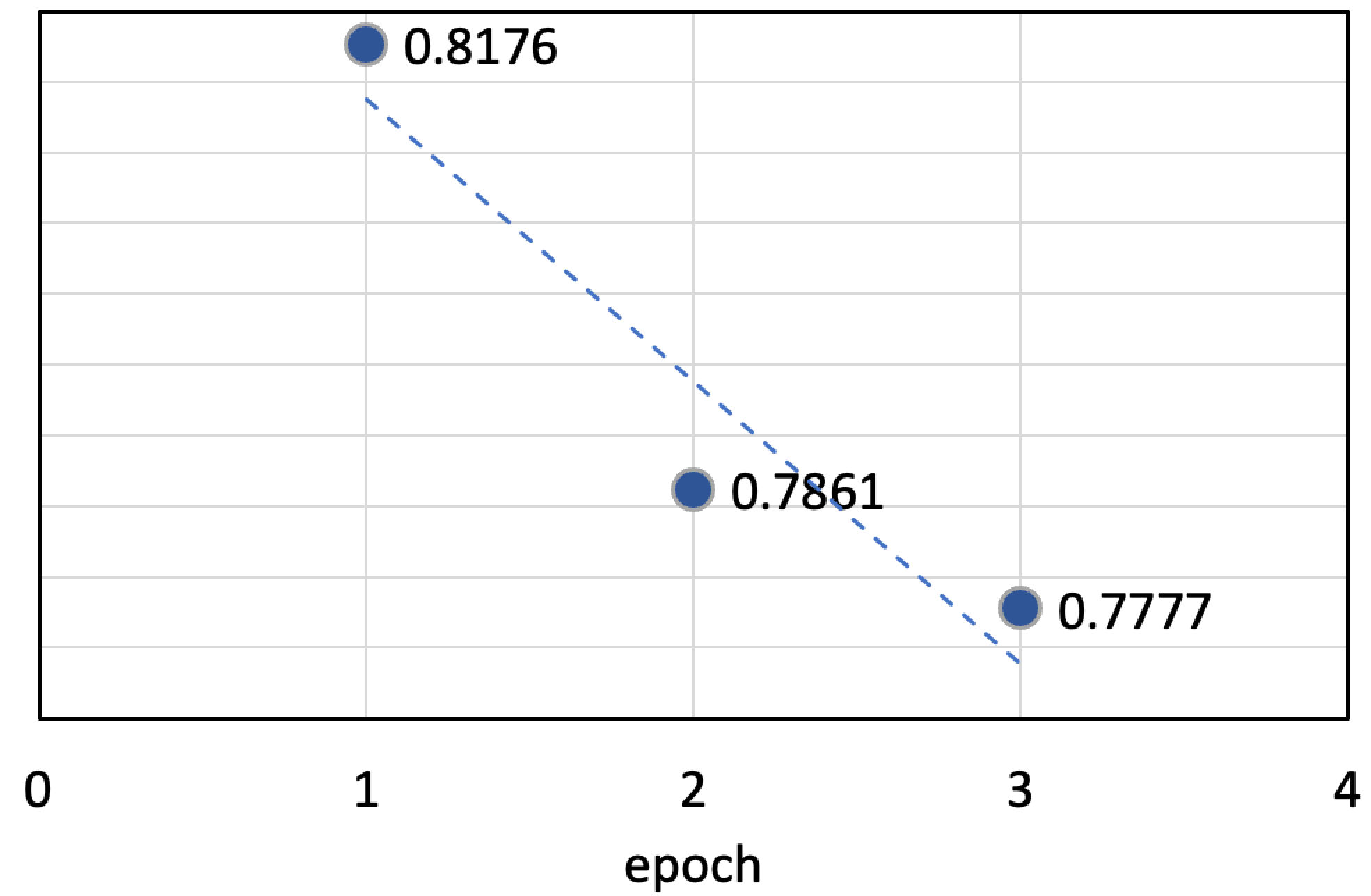}
        \caption{Epoch} 
        \label{fig:epoch}
    \end{subfigure}
    \hfill
    \begin{subfigure}[t]{0.45\textwidth}
        \centering
        \includegraphics[width=\linewidth]{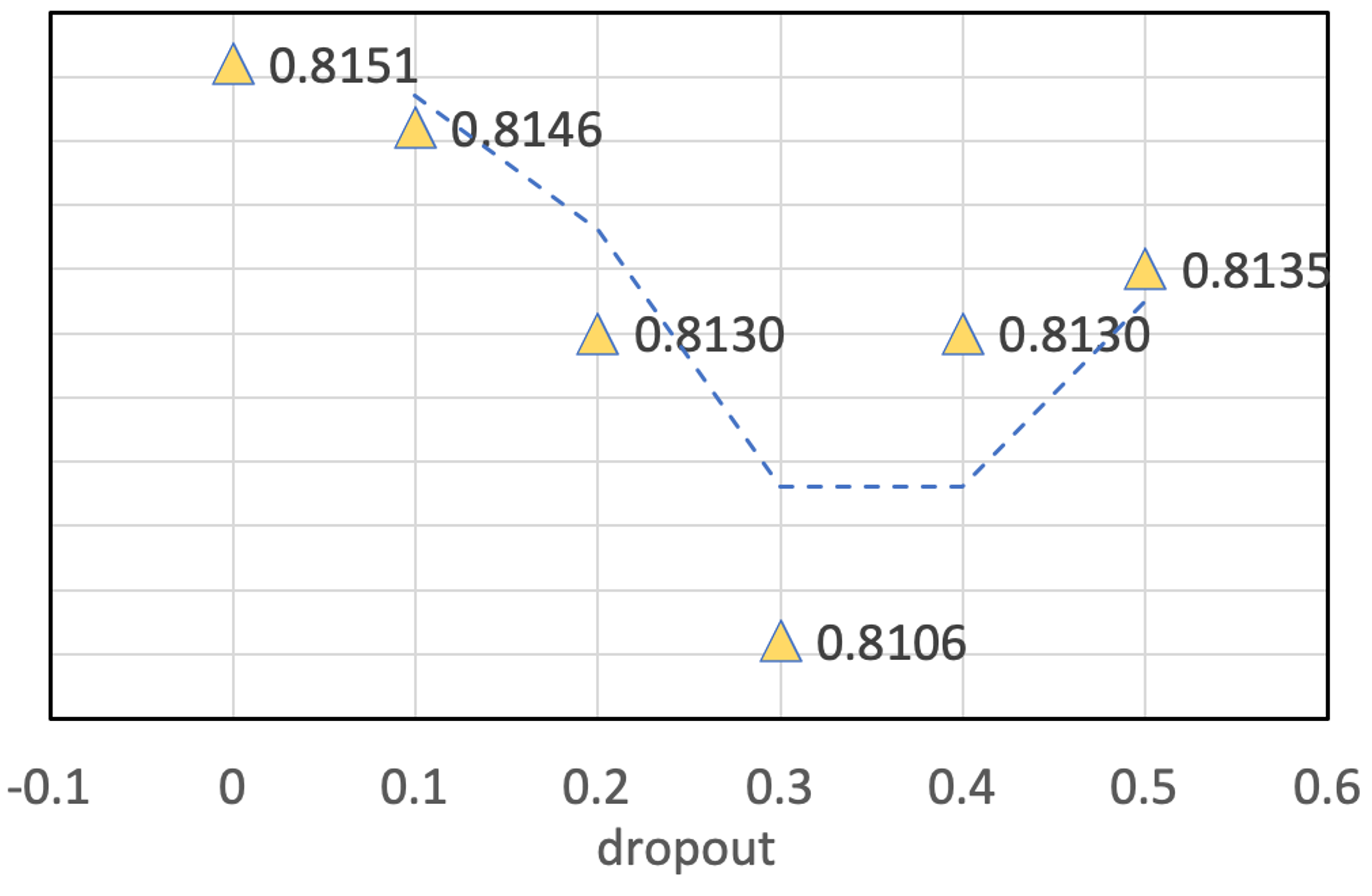}
        \caption{Dropout}
        \label{fig:dropout}
    \end{subfigure}
    
    \vspace{0.3cm}
    
    \begin{subfigure}[t]{0.45\textwidth}
        \centering
        \includegraphics[width=\linewidth]{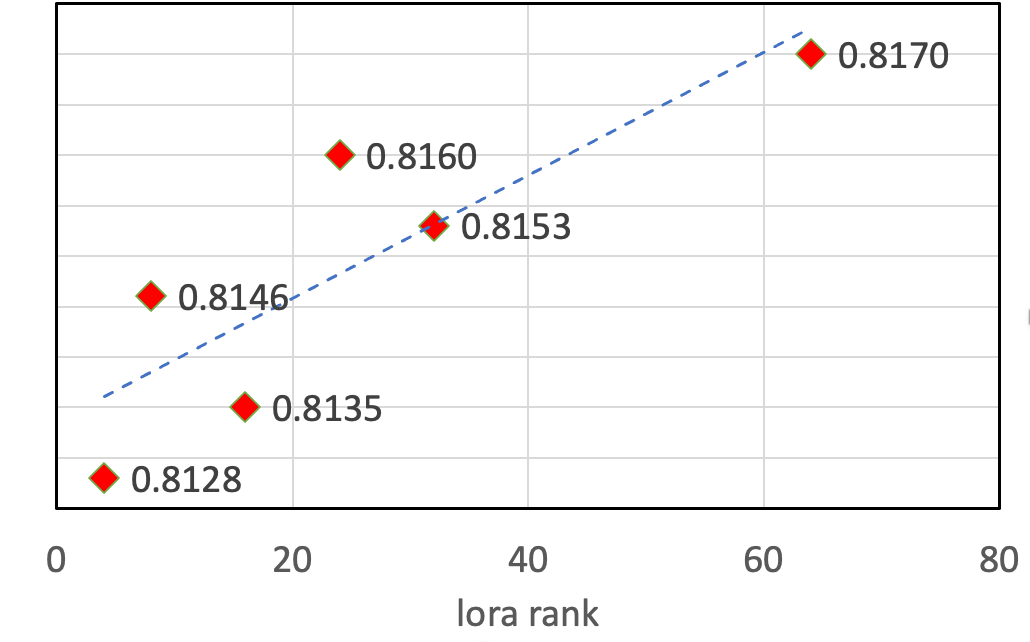}
        \caption{LoRA Rank}
        \label{fig:rank}
    \end{subfigure}

    \caption{The Impact of Training Settings on asLLR.}
    \label{fig:params}
\end{figure}

Although we did not observe a significantly evident scaling law, based on experimental evidence, we can reasonably conclude that improvements in the technical foundation positively affect the performance of the asLLR technology. This suggests that, compared to traditional CTR models, clue ranking technologies based on LLM techniques could be a more promising direction.

\begin{figure}[h]
\centering
\centerline{\includegraphics[width=0.7\linewidth]{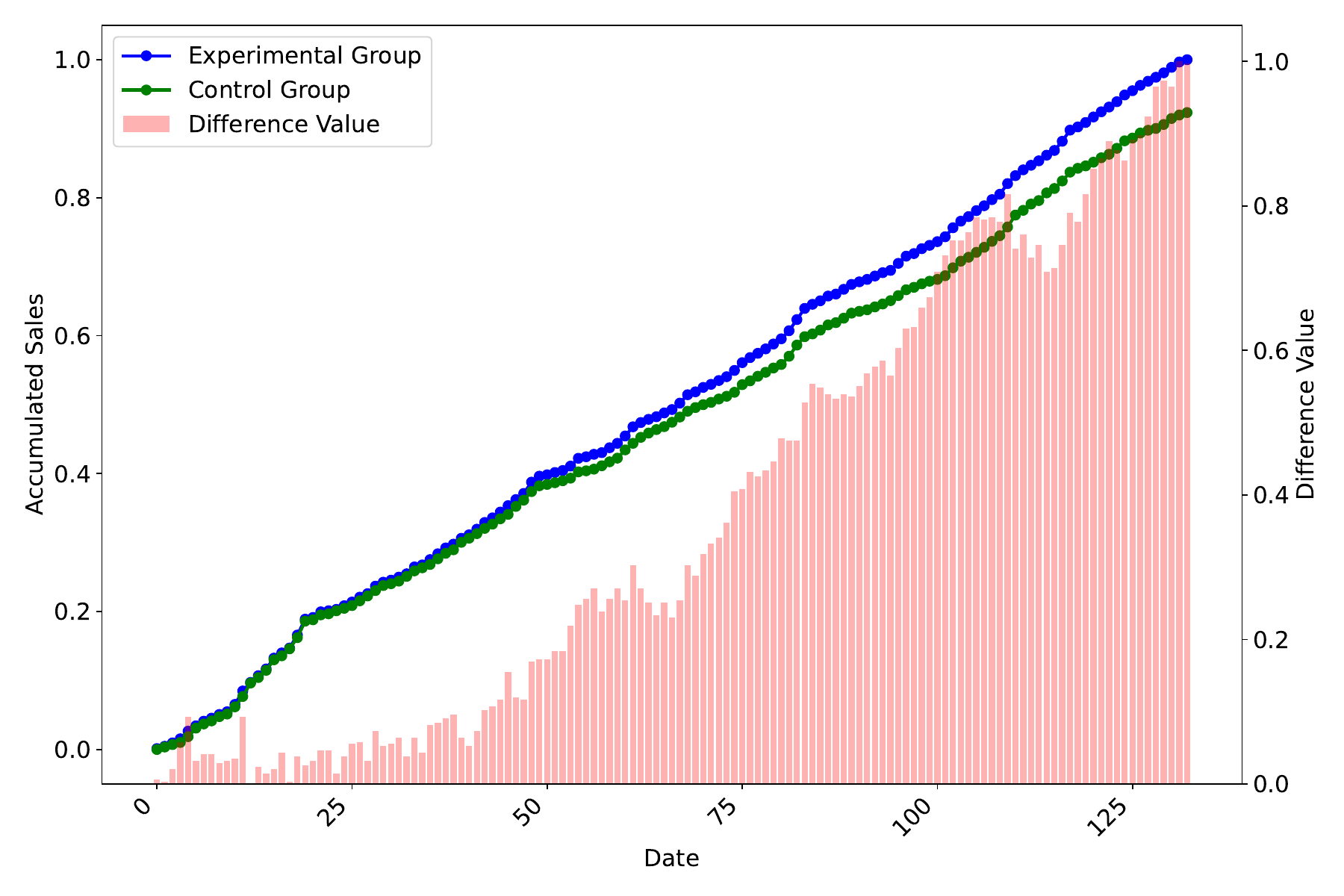}}
\caption{\quad  The online experimental results of asLLR. The specific values have been normalized due to legal risk considerations.}
\label{fig:result}
\end{figure}

\subsection{Parameter Optimization}

Through our experiments, we evaluated the impact on asLLR model performance by varying the number of training epochs, dropout rates, and LoRA rank. As shown in Figure \ref{fig:epoch}, we found that training for a single epoch yielded the best results. This suggests that fine-tuning a large language model is still highly prone to overfitting, and addressing LLM overfitting remains a critical area for further research. In Figure \ref{fig:dropout},
the performance based on dropout rates appeared to exhibit a U-shaped pattern, with better results at both ends and poorer performance in the middle. We recommend setting the dropout rate to 0.5. Finally, we observed that an increase in LoRA rank, which allows for more trainable parameters, generally led to improved model performance. 
As shown in Figure \ref{fig:rank}, this aligns with our intuition; however, it is important to note that excessively large LoRA ranks might also introduce a risk of overfitting, based on our experimental observations.

\subsection{Online Performance Validation}
We also validate the asLLR model
through a rigorous experimental grouping method. In the real online A/B testing scenario, the sales specialists were divided into two groups with equivalent baselines: a control group using traditional CTR techniques for lead evaluation and an experimental group employing asLLR techniques. Specialists prioritized following up on high-quality leads according to the lead quality ranking. After almost five months of implementation, we evaluated the performance of both groups. Results indicate that the experimental group experienced a $9.5\%$ increase in lead conversion by specialists compared to the control group. The process of the overall experiment in a specific province, as well as the lead conversion and difference between the experimental and control groups, is illustrated in Figure \ref{fig:result}. The significant improvement in online performance not only supports the validity of the technical metrics used for evaluation but also suggests that this technology may possess substantial commercial value.

\subsection{Research on Extremely Long Texts Features }
In real-world evaluation, we found that the original call-record text exhibited low information density. Therefore, we processed all call records using a text summarization model, resulting in a nearly 60\% reduction in the token length of the overall call records. Table \ref{Dif context} displays the distribution of text sequence lengths before and after summarization.
\begin{table}[ht]
\centering
\small
\caption{Distribution of Text Sequence Lengths Before and After Summarization.}
\label{Dif context}
\resizebox{0.8\columnwidth}{!}{
\begin{tabular}{ccccccccccc}
\hline
Quantile & 0.1 & 0.2 & 0.3 & 0.4 & 0.5 & 0.6 & 0.7 & 0.8 & 0.9 & 0.95 \\ \hline
Original & 0 & 0 & 255 & 457 & 734 & 1138 & 1756 & 2868 & 5202 & 8727 \\ \hline
Summarized & 0 & 0 & 151 & 260 & 398 & 579 & 837 & 1247 & 2054 & 3007 \\ \hline
\end{tabular}
}
\end{table}

\begin{figure}[htbp]
\centering
\centerline{\includegraphics[width=0.7\linewidth]{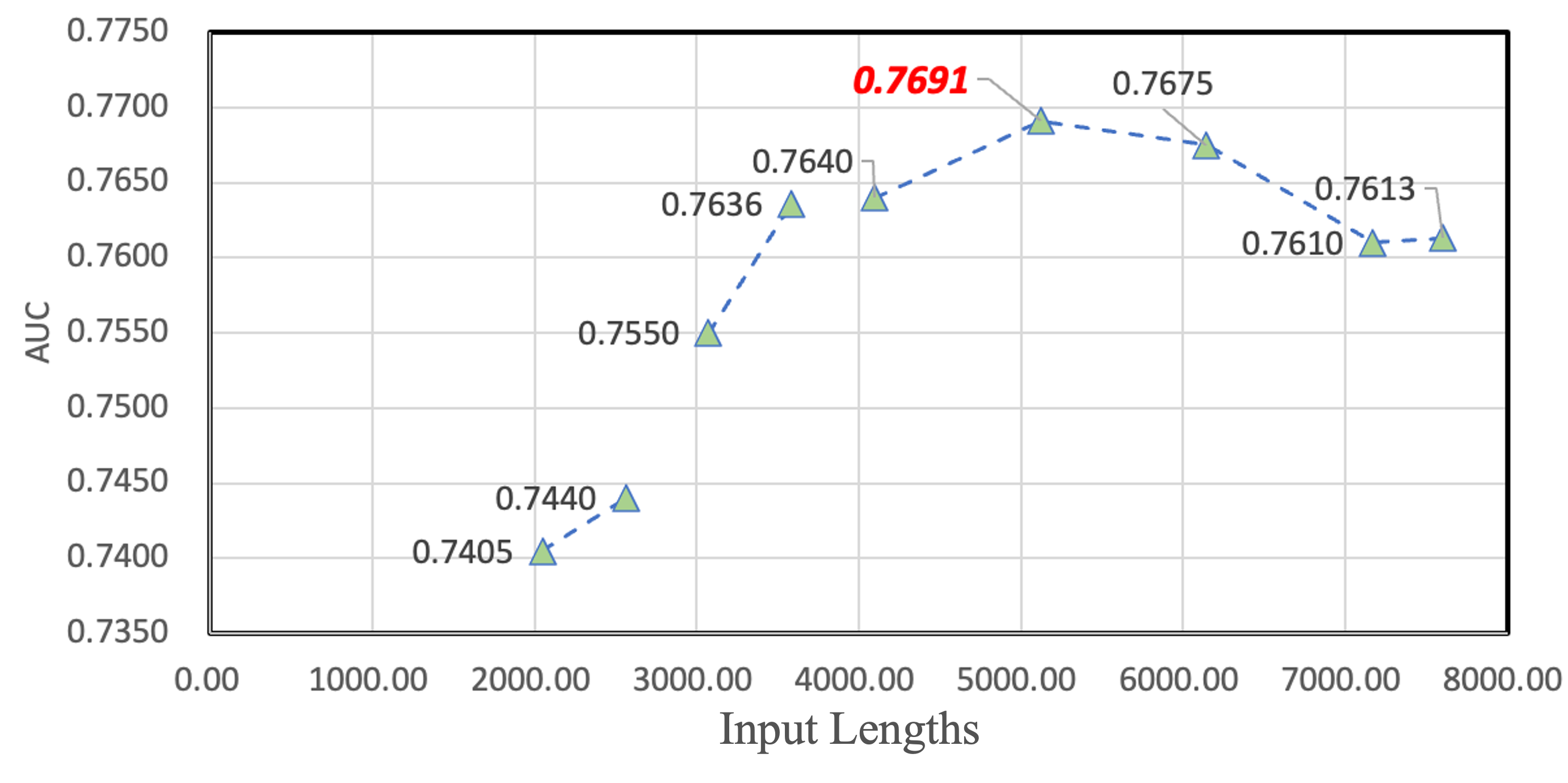}}
\caption{Performance AUC of the asLLR Model with Varying Context Input Lengths.}
\label{diff input}
\end{figure}

\subsubsection{\textbf{Impact of Input Context Length.}} 
To investigate the performance issues faced by the asLLR model when dealing with varying context input lengths, we designed a series of experiments. The results, depicted in Figure \ref{diff input}, demonstrate that as the context length increases, the technical performance of the asLLR model initially improves, reaching a peak when the context length approaches 5000 tokens, and then begins to decline. As shown in Table \ref{Dif context}, over 10\% of the test samples have a maximum length exceeding 5000 tokens. Theoretically, as the model incorporates longer contexts, the amount of input information increases, potentially improving the model's technical metrics. However, when the context length continues to increase, the model’s performance does not further improve with the enriched modeling of information; rather, the technical metrics start to decline. This suggests that the model might struggle to effectively capture critical information embedded within long contexts that are high in noise.

\subsubsection{\textbf{Text Summarization.}} 
\begin{figure}
    \centering
    \includegraphics[width=0.7\linewidth]{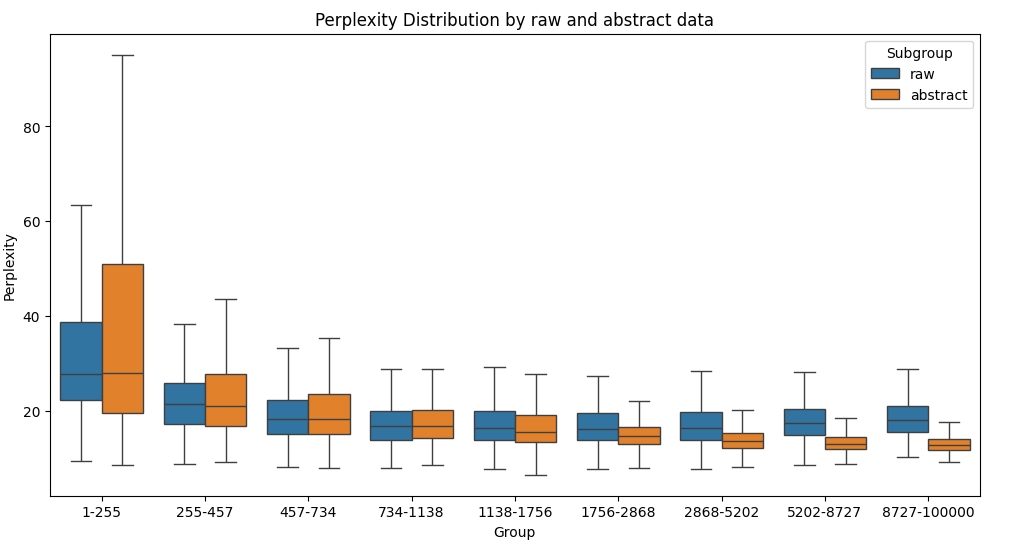}
    \caption{The Impact of Text Summarization Techniques on the Perplexity of Contexts with Different Lengths.}
    \label{text diffusion}
\end{figure}

\begin{figure}[ht]
\centering
\centerline{\includegraphics[width=0.7\linewidth]{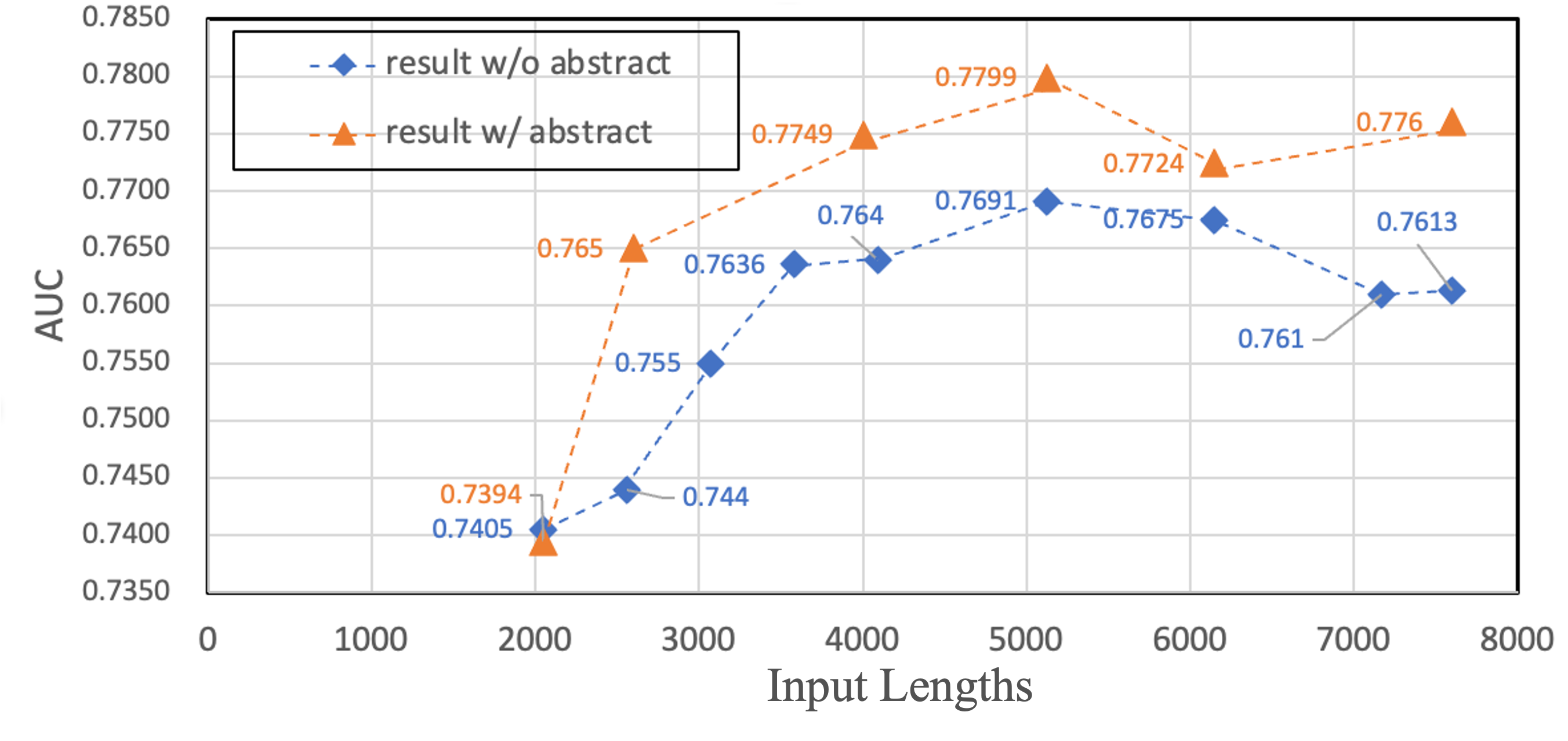}}
\caption{
A Comparative Analysis of the Performance between \textcolor{orange}{asLLR w/ abstract} and \textcolor{blue}{asLLR w/o  abstract}  across Various Context Length Inputs.}
\label{dub}
\end{figure}
To tackle the aforementioned issue, we postulated that high input perplexity could be a contributing factor. Consequently, we calculated perplexity for texts of various lengths alongside their summaries. As depicted in Figure \ref{text diffusion}, the results suggest a correlation between the length of the original input and its perplexity. We investigated whether certain technical strategies could mitigate input perplexity and enhance information density. Leveraging text summarization techniques grounded in large language models (LLMs), we summarized the input dialogues and compared the perplexity of the resultant sentences. Detailed results in Figure \ref{dub} highlight that summarization techniques effectively lower input perplexity in long-context scenarios.

Subsequently, we integrate summarized text into the model to assess the effect of incorporating versus omitting summaries, maintaining consistent experimental parameters throughout. As illustrated in Figure \ref{dub}, the findings reveal that the asLLR model, when augmented with summarization strategies, exhibits notable enhancement in processing lengthy textual inputs, evidenced by an augmentation in the average AUC from 0.7584 to 0.7679. This performance augmentation was consistently observed, with the exception of scenarios involving shorter input contexts. In such scenarios, the application of summarization techniques may inadvertently escalate input perplexity, thereby leading to diminished model efficacy.

\section{Conclusions}
In this study, we present asLLR, an innovative lead ranking model based on large language models (LLMs) and employing a decoder-only architecture. By simultaneously modeling and integrating both tabular and textual features, asLLR enhances the efficacy of automotive sales lead ranking. Given the current scarcity of research and the lack of publicly available datasets in the academic domain for the automotive sales sector, we independently developed a dataset for lead quality evaluation, derived from real-world customer data from a specific electric vehicle brand. This dataset was instrumental in evaluating the performance of conventional CTR prediction models in comparison to the asLLR model. To address the practical challenges posed by excessively long text feature inputs, which can complicate model training, we integrated a text summarization module. This module performs text summarization and knowledge compression on the input textual features, thereby enhancing the model's performance in scenarios that involve extended text feature inputs.

\bibliographystyle{unsrt}  
\bibliography{references}

\begin{thebibliography}{10}

\bibitem{fischer2022artificial}
Heiko Fischer, Sven Seidenstricker, Thomas Berger, and Timo Holopainen.
\newblock Artificial intelligence in b2b sales: Impact on the sales process.
\newblock {\em Artificial Intelligence and Social Computing}, 28(28):135--142, 2022.

\bibitem{alavi2024salesperson}
Sascha Alavi, Johannes Habel, and Arnd Vomberg.
\newblock Salesperson lifecycle management: Challenges and research priorities, 2024.

\bibitem{fehrenbach2025artificial}
David Fehrenbach, Carolina Herrando, and Benjamin {\"O}sterle.
\newblock Artificial intelligence applications in the b2b sales funnel.
\newblock {\em Journal of Business-to-Business Marketing}, pages 1--24, 2025.

\bibitem{sf}
Kung-Hsiang Huang, Akshara Prabhakar, Sidharth Dhawan, Yixin Mao, Huan Wang, Silvio Savarese, Caiming Xiong, Philippe Laban, and Chien-Sheng Wu.
\newblock Crmarena: Understanding the capacity of llm agents to perform professional crm tasks in realistic environments, 2025.

\bibitem{zhang2016deep}
Weinan Zhang, Tianming Du, and Jun Wang.
\newblock Deep learning over multi-field categorical data: --a case study on user response prediction.
\newblock In {\em Advances in Information Retrieval: 38th European Conference on IR Research, ECIR 2016, Padua, Italy, March 20--23, 2016. Proceedings 38}, pages 45--57. Springer, 2016.

\bibitem{zhang2024wukong}
Buyun Zhang, Liang Luo, Yuxin Chen, Jade Nie, Xi~Liu, Daifeng Guo, Yanli Zhao, Shen Li, Yuchen Hao, Yantao Yao, et~al.
\newblock Wukong: Towards a scaling law for large-scale recommendation.
\newblock {\em arXiv preprint arXiv:2403.02545}, 2024.

\bibitem{lmaskb}
Badr AlKhamissi, Millicent Li, Asli Celikyilmaz, Mona Diab, and Marjan Ghazvininejad.
\newblock A review on language models as knowledge bases, 2022.

\bibitem{gpt3analysis}
Luciano Floridi and Massimo Chiriatti.
\newblock Gpt-3: Its nature, scope, limits, and consequences.
\newblock {\em Minds and Machines}, 30:681--694, 2020.

\bibitem{agent}
Zhiheng Xi, Wenxiang Chen, Xin Guo, Wei He, Yiwen Ding, Boyang Hong, Ming Zhang, Junzhe Wang, Senjie Jin, Enyu Zhou, Rui Zheng, Xiaoran Fan, Xiao Wang, Limao Xiong, Yuhao Zhou, Weiran Wang, Changhao Jiang, Yicheng Zou, Xiangyang Liu, Zhangyue Yin, Shihan Dou, Rongxiang Weng, Wensen Cheng, Qi~Zhang, Wenjuan Qin, Yongyan Zheng, Xipeng Qiu, Xuanjing Huang, and Tao Gui.
\newblock The rise and potential of large language model based agents: A survey, 2023.

\bibitem{cheng2016wide}
Heng-Tze Cheng, Levent Koc, Jeremiah Harmsen, Tal Shaked, Tushar Chandra, Hrishi Aradhye, Glen Anderson, Greg Corrado, Wei Chai, Mustafa Ispir, et~al.
\newblock Wide \& deep learning for recommender systems.
\newblock In {\em Proceedings of the 1st workshop on deep learning for recommender systems}, pages 7--10, 2016.

\bibitem{deepfm}
Huifeng Guo, Ruiming Tang, Yunming Ye, Zhenguo Li, and Xiuqiang He.
\newblock Deepfm: A factorization-machine based neural network for ctr prediction, 2017.

\bibitem{dcn}
Ruoxi Wang, Bin Fu, Gang Fu, and Mingliang Wang.
\newblock Deep \& cross network for ad click predictions, 2017.

\bibitem{dcnm}
Ruoxi Wang, Rakesh Shivanna, Derek~Zhiyuan Cheng, Sagar Jain, Dong Lin, Lichan Hong, and Ed~H. Chi.
\newblock Dcn-m: Improved deep \& cross network for feature cross learning in web-scale learning to rank systems.
\newblock {\em ArXiv}, abs/2008.13535, 2020.

\bibitem{dcnv2}
Ruoxi Wang, Rakesh Shivanna, Derek Cheng, Sagar Jain, Dong Lin, Lichan Hong, and Ed~Chi.
\newblock Dcn v2: Improved deep \&amp; cross network and practical lessons for web-scale learning to rank systems.
\newblock In {\em Proceedings of the Web Conference 2021}, WWW ’21. ACM, April 2021.

\bibitem{graphfm}
Shu Wu, Zekun Li, Yunyue Su, Zeyu Cui, Xiaoyu Zhang, and Liang Wang.
\newblock Graphfm: Graph factorization machines for feature interaction modeling, 2024.

\bibitem{hirs}
Yixin Su, Yunxiang Zhao, Sarah~Monazam Erfani, Junhao Gan, and Rui Zhang.
\newblock Detecting arbitrary order beneficial feature interactions for recommender systems.
\newblock {\em Proceedings of the 28th ACM SIGKDD Conference on Knowledge Discovery and Data Mining}, 2022.

\bibitem{el2021automatic}
Wafaa~S El-Kassas, Cherif~R Salama, Ahmed~A Rafea, and Hoda~K Mohamed.
\newblock Automatic text summarization: A comprehensive survey.
\newblock {\em Expert systems with applications}, 165:113679, 2021.

\bibitem{nallapati2016abstractive}
Ramesh Nallapati, Bowen Zhou, Caglar Gulcehre, Bing Xiang, et~al.
\newblock Abstractive text summarization using sequence-to-sequence rnns and beyond.
\newblock {\em arXiv preprint arXiv:1602.06023}, 2016.

\bibitem{shi2021neural}
Tian Shi, Yaser Keneshloo, Naren Ramakrishnan, and Chandan~K Reddy.
\newblock Neural abstractive text summarization with sequence-to-sequence models.
\newblock {\em ACM Transactions on Data Science}, 2(1):1--37, 2021.

\bibitem{fabbri2019multi}
Alexander~R Fabbri, Irene Li, Tianwei She, Suyi Li, and Dragomir~R Radev.
\newblock Multi-news: A large-scale multi-document summarization dataset and abstractive hierarchical model.
\newblock {\em arXiv preprint arXiv:1906.01749}, 2019.

\bibitem{li2019abstractive}
Wei Li and Hai Zhuge.
\newblock Abstractive multi-document summarization based on semantic link network.
\newblock {\em IEEE Transactions on Knowledge and Data Engineering}, 33(1):43--54, 2019.

\bibitem{deyoung2021ms2}
Jay DeYoung, Iz~Beltagy, Madeleine van Zuylen, Bailey Kuehl, and Lucy~Lu Wang.
\newblock Ms2: Multi-document summarization of medical studies.
\newblock {\em arXiv preprint arXiv:2104.06486}, 2021.

\bibitem{nallapati2017summarunner}
Ramesh Nallapati, Feifei Zhai, and Bowen Zhou.
\newblock Summarunner: A recurrent neural network based sequence model for extractive summarization of documents.
\newblock In {\em Proceedings of the AAAI conference on artificial intelligence}, volume~31, 2017.

\bibitem{narayan2018ranking}
Shashi Narayan, Shay~B Cohen, and Mirella Lapata.
\newblock Ranking sentences for extractive summarization with reinforcement learning.
\newblock {\em arXiv preprint arXiv:1802.08636}, 2018.

\bibitem{zhong2020extractive}
Ming Zhong, Pengfei Liu, Yiran Chen, Danqing Wang, Xipeng Qiu, and Xuanjing Huang.
\newblock Extractive summarization as text matching.
\newblock {\em arXiv preprint arXiv:2004.08795}, 2020.

\bibitem{zhang2020pegasus}
Jingqing Zhang, Yao Zhao, Mohammad Saleh, and Peter Liu.
\newblock Pegasus: Pre-training with extracted gap-sentences for abstractive summarization.
\newblock In {\em International conference on machine learning}, pages 11328--11339. PMLR, 2020.

\bibitem{mihalcea2004textrank}
Rada Mihalcea and Paul Tarau.
\newblock Textrank: Bringing order into text.
\newblock In {\em Proceedings of the 2004 conference on empirical methods in natural language processing}, pages 404--411, 2004.

\bibitem{devlin2018bert}
Jacob Devlin.
\newblock Bert: Pre-training of deep bidirectional transformers for language understanding.
\newblock {\em arXiv preprint arXiv:1810.04805}, 2018.

\bibitem{pu2023summarization}
Xiao Pu, Mingqi Gao, and Xiaojun Wan.
\newblock Summarization is (almost) dead.
\newblock {\em arXiv preprint arXiv:2309.09558}, 2023.

\bibitem{inproceedings}
Richardson, Matthew, Dominowska, Ewa, Ragno, and Robert.
\newblock Predicting clicks: Estimating the click-through rate for new ads.
\newblock 05 2007.

\bibitem{hu2022lora}
Edward~J Hu, Yelong Shen, Phillip Wallis, Zeyuan Allen-Zhu, Yuanzhi Li, Shean Wang, Lu~Wang, and Weizhu Chen.
\newblock Lo{RA}: Low-rank adaptation of large language models.
\newblock In {\em International Conference on Learning Representations}, 2022.

\bibitem{lindell2021autoint}
David~B Lindell, Julien~NP Martel, and Gordon Wetzstein.
\newblock Autoint: Automatic integration for fast neural volume rendering.
\newblock In {\em Proceedings of the IEEE/CVF Conference on Computer Vision and Pattern Recognition}, pages 14556--14565, 2021.

\bibitem{kaplan2020scaling}
Jared Kaplan, Sam McCandlish, Tom Henighan, Tom~B Brown, Benjamin Chess, Rewon Child, Scott Gray, Alec Radford, Jeffrey Wu, and Dario Amodei.
\newblock Scaling laws for neural language models.
\newblock {\em arXiv preprint arXiv:2001.08361}, 2020.

\bibitem{kirkpatrick2017overcoming}
James Kirkpatrick, Razvan Pascanu, Neil Rabinowitz, Joel Veness, Guillaume Desjardins, Andrei~A Rusu, Kieran Milan, John Quan, Tiago Ramalho, Agnieszka Grabska-Barwinska, et~al.
\newblock Overcoming catastrophic forgetting in neural networks.
\newblock {\em Proceedings of the national academy of sciences}, 114(13):3521--3526, 2017.

\end{thebibliography}

\end{document}